\documentclass[letterpaper, inpress]{jds} 

\jdsmonth{March}                 
\jdsyear{2021}                  
\jdsvolume{xx}                  
\jdsissue{xx}                   
\jdsdoi{10.6339/21-JDS1010}      
\jdsreceived{December, 2020}       
\jdsaccepted{February, 2021}         

\usepackage{amsfonts,amsmath,amssymb,amsthm,gensymb}
\usepackage{booktabs}

\usepackage{lipsum}

\title[Air Quality and COVID-19]{Investigating the Relationship Between Air Quality and COVID-19 Transmission}

\author[1]{Laura Albrecht\footnote{All authors contributed equally to this work}}
\author[2]{Paulina Czarnecki\thanks{Corresponding author. Email: pc2943@columbia.edu}}
\author[3]{Bennet Sakelaris}
\affil[1]{Department of Applied Mathematics and Statistics, Colorado School of Mines, Golden, CO, USA}
\affil[2]{Department of Applied Physics and Applied Mathematics, Columbia University, New York City, NY, USA}
\affil[3]{Department of Engineering Sciences and Applied Mathematics, Northwestern University, Evanston, IL, USA}

\begin{document}

\maketitle

\begin{abstract}
  It is hypothesized that short-term exposure to air pollution may influence the transmission of aerosolized pathogens such as COVID-19. We used data from 23 provinces in Italy to build a generalized additive model to investigate the association between the effective reproductive number of the disease and air quality while controlling for ambient environmental variables and changes in human mobility. The model finds that there is a positive, nonlinear relationship between the density of particulate matter in the air and COVID-19 transmission, which is in alignment with similar studies on other respiratory illnesses.\@
\end{abstract}


\section{Introduction}%
\label{sec:intro}

\noindent
Poor air quality is a well-documented public health concern, causing respiratory illnesses, negatively impacting cardiovascular health, and decreasing life expectancy \citep{delfino2005potential, krewski2009evaluating, lelieveld2015contribution, janssen2013short}. Especially harmful is fine particulate matter (with diameter less than 2.5~$\mu$m, called PM$_{2.5}$) suspended in the air   \citep{delfino2005potential, laden2000association, schwartz1996daily}. These particles, which are mainly created by industrial combustion processes and atmospheric reactions, can deeply penetrate the lungs and cause damage to the respiratory system over time \citep{xing2016impact, tucker2000overview}. Furthermore, it is hypothesized that viruses may attach themselves to these suspended particles and infect hosts upon inhalation \citep{ciencewicki2007air, liang2014pm, contini2020does, jiang2020effect}. This effect is particularly pronounced in cities with extreme levels of air pollution, such as Beijing, where cases of influenza have been shown to increase with the density of PM$_{2.5}$ in the air \citep{ciencewicki2007air, liang2014pm, feng2016impact, su2019short, jiang2020effect}. \newline
\indent Since December 2019, when the novel coronavirus SARS-CoV-2 was first identified in Wuhan, China, it has erupted into a global pandemic \citep{zhu2020novel,jiang2020effect, yongjian2020association}. Many risk factors and comorbidities have been found, including obesity, pre-existing health conditions such as diabetes or hypertension, and advanced age \citep{richardson2020presenting, yang2020prevalence}. An increase in environmental factors such as air pollution, temperature, and humidity have also been found to be associated with increased COVID-19 case counts early in the pandemic in China \citep{yongjian2020association, jiang2020effect}, however the effects of PM$_{2.5}$ remain largely unexplored as the pandemic continues to develop. Furthermore, recent events such as raging wildfires in the western United States have increased concentrations of PM$_{2.5}$ in the air \citep{o2019contribution, jaffe2008interannual}; it would be useful to understand whether these changes can affect the spread of COVID-19.
\newline \indent In this paper, we analyze the effective reproductive number ($R_t$) of the disease, which is the average number of secondary infections caused by an infectious individual on day $t$. Because of this time dependency, $R_t$ acts as a measurement of daily transmission intensity and as a result, is frequently utilized for decision-making during a pandemic \citep{why_rt}. After using the open-source software rt.live \citep{rtlive2020} to calculate $R_t$ in 23 provinces in Italy, we studied its association with PM$_{2.5}$ using a generalized additive models (GAM). In the GAM, we also controlled for other environmental variables, including humidity, temperature, and changes in mobility due to lockdown. We found that at concentrations of PM$_{2.5}$ greater than about 70~$\mu$g/m$^3$, poor air quality is positively correlated with increased $R_t$ of COVID-19. Furthermore, we performed a 23-fold cross-validation procedure in order to test the sensitivity of the model on each province and test for selection bias. We found that our model was robust against this test.

\section{Methods}%
\label{sec:Methods}
\subsection{Study Area} 
\begin{figure}
    \centering
    \includegraphics[scale=0.2]{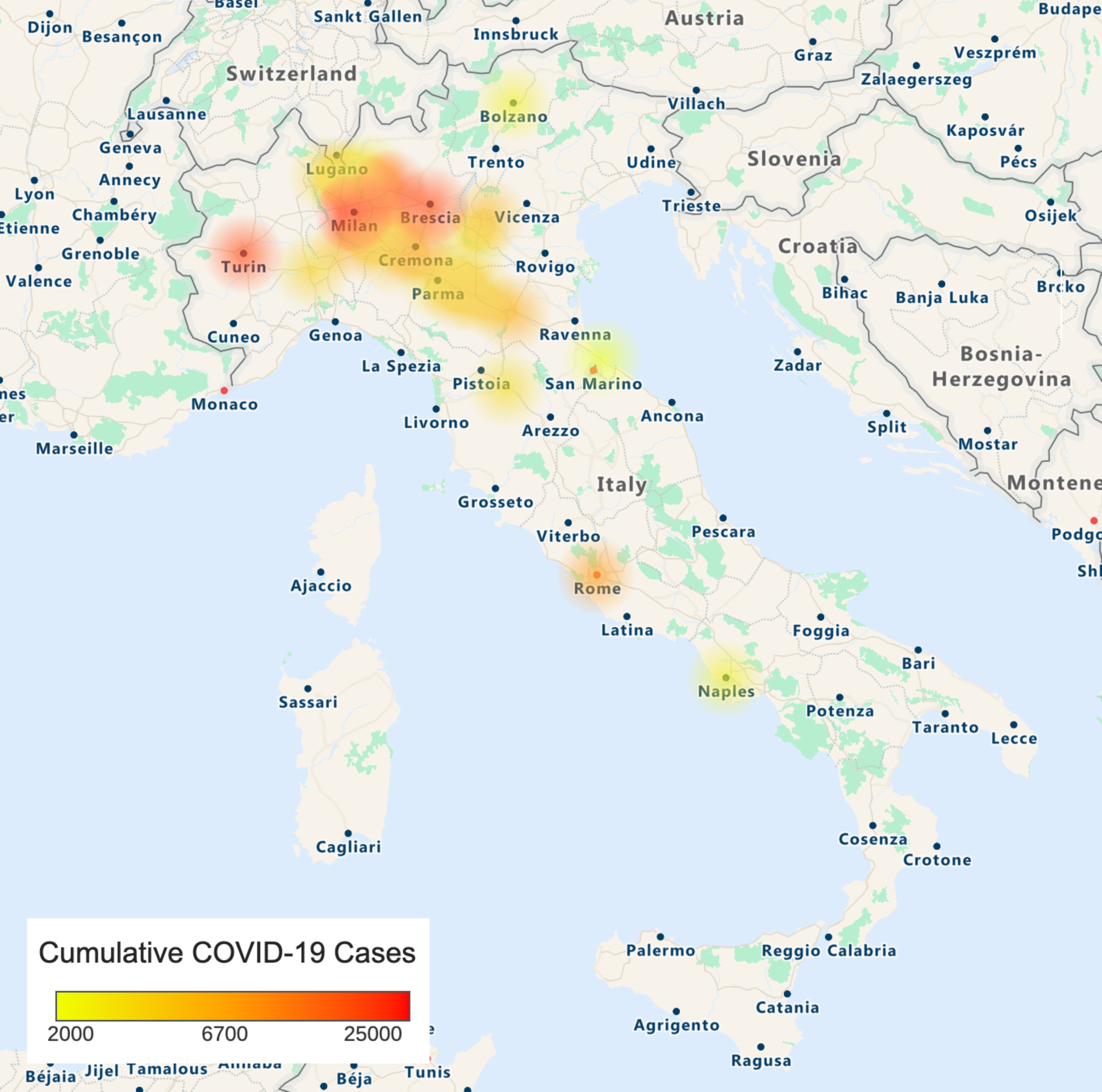}
    \caption{A geospatial heat map with locations of the 23 provinces included in the study, colored by the cumulative number of COVID-19 cases as of August 1, 2020.}
    \label{fig:map}
\end{figure}

In order to ensure that we had sufficient data to extract meaningful results, we considered only Italian provinces that reported more than 2000 cumulative cases of COVID-19 during the study period of February 24, 2020 to August 1, 2020. We then removed locations where environmental data was unavailable, resulting in the final study area of the 23 provinces shown in Figure \ref{fig:map}. These provinces accounted for over 63\% of total Italy’s COVID-19 cases over the time frame.

\subsection{Data} 
The environmental variables included in the study were daily median temperature, relative humidity, and PM$_{2.5}$. Temperature and relative humidity were collected from the NOAA Integrated Surface Database using the \pkg{worldmet} package in \proglang{R} \citep{worldmet}, while PM$_{2.5}$ data was taken from the World Air Quality Index (https://aqicn.org/).

We also controlled for the effects of lockdown by incorporating data from province-specific Google Community Mobility Reports (https://google.com/covid19/mobility/),
which quantify the daily change in aggregate mobility compared to a pre-pandemic baseline. The dataset was generated using anonymized data from Google Maps to measure daily visitors in six categories of locations: residential areas, retail locations, transit stations, grocery stores, parks, and workplaces. In order to gain perspective on how mobility has changed due to the pandemic, these measures were divided by a median value of visitors for that location for that day of the week, which was measured prior the onset of the pandemic. For this study, we took our mobility variable to be the change in mobility in residential areas; however, we found similar results when we substituted data from any of the other five categories of locations provided in the Google Community Mobility Reports.

Daily tests counts and confirmed cases were collected directly from the Italian Department of Civil Protection (https://github.com/pcm-dpc/COVID-19) and were used to estimate the value of the effective reproductive number, $R_t$.

$R_t$ was estimated using an open source model from \cite{rtlive2020} called rt.live, which we adapted to be compatible with our data. Notably, the model estimates total cases by scaling confirmed cases according to testing volume. It also approximates a time series of cases by day of infection using an onset delay distribution which was empirically driven by data from  \cite{onset_delay} while assuming a five day incubation period as evidenced by work from \cite{incubation}. Finally, it uses the generation time derived by \cite{generation_time} with mean 4.7 days and standard deviation 2.9 days to estimate $R_t$.

\begin{figure}[t]
    \centering
    \includegraphics[width=\linewidth]{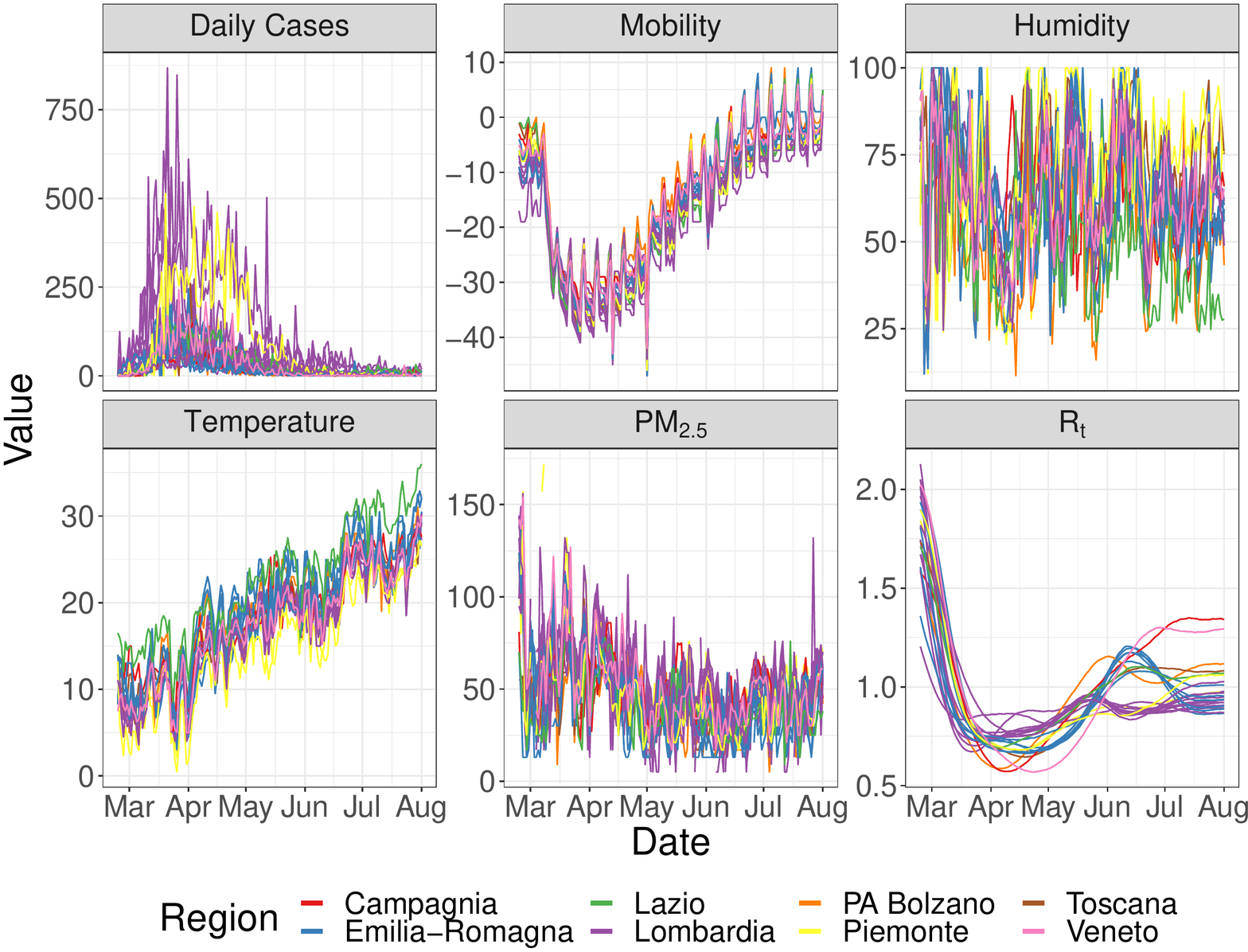}
    \caption{The time series of daily cases, temperature, humidity, and PM$_{2.5}$ in 23 Italian provinces. Each line represents one province and the colors are grouped by region.}
    \label{fig:environ}
\end{figure}

The time series of new COVID-19 cases, mobility, PM$_{2.5}$, median temperature, median humidity, and $R_t$ from February 24, 2020 to August 1st, 2020 are shown in Figure \ref{fig:environ}. The peak in cases occurred between March 14th and April 18th in each province then cases steadily decreased. The effective reproductive number however decreases until mid-April and starts to increase again in most provinces over the summer. A seasonal increase in temperature is present as the data was collected from late winter to mid-summer, while humidity and PM$_{2.5}$ do not follow any obvious trends. Descriptive statistics of our data are available in Table \ref{Tab:data}.
\begin{table}
\centering

\begin{tabular}{lrrrr}
\hline
\hline
                       & Mean  & SD    & Min   & Max   \\ 
\hline
\hline
Daily Cases                        & 42.29 & 76.93 & 0     & 868   \\
Daily Tests                        & 968.93 & 1136.27 & 0 & 18256  \\
Temperature ($^o$C)                & 18.06 & 6.60   & 0.50  & 36   \\
Relative Humidity (\%)             & 63.84 & 17.67 & 11.50 & 100   \\
\text{PM}$_{2.5}$ ($\mu$g/m$^3$)   & 48.52 & 21.67 & 5     & 172   \\
Mobility Decrease (\%)             & 15.57 & 12.31 & -9    & 47    \\
R$_t$                              & 0.94  & 0.24  & 0.57  & 2.12  \\
\hline
\end{tabular}
\caption{Mean, standard deviation, minimum, and maximum of each variable considered across all cities and days, including the estimated value of R$_t$.}
\label{Tab:data}
\end{table}

\subsection{Statistical Model} 
Following other studies on the effects of ambient environmental variables on viral transmission \citep{air_pollution_and_mortality, dengue_gam, temperature_covid_gam_brazil, temperature_covid_gam_china, leishmaniasis_gam, feng2016impact}, we used a generalized additive model (GAM) to analyze the COVID-19 pandemic. The GAM is a semi-parametric extension of the generalized linear model in which the response variable (relative transmission rate) relies on a linear combination of smooth functions of the predictor variables (humidity, PM$_{2.5}$, temperature, and mobility). Our GAM was defined using a Gaussian distribution and the following equation:
\begin{equation}
\label{eqn:gam}
    \log(R_{t,i}) = s(\text{Mobility}_{t,i}) + s(\text{Temp}_{t,i}) + s(\text{Hum}_{t,i}) + s(\text{PM}_{2.5_{t,i}}) + \lambda_i,
\end{equation}
where $t$ denotes the date, $i$ denotes the province, and $s(\cdot)$ denotes the thin plate basis spline used to smooth the data. Parameters of the splines were estimated using the GCV method, and the splines were constrained to have at most five degrees of freedom. This constraint was imposed to minimize overfitting while accurately describing the trends of the data. The results of the model were robust to changes in the degrees of freedom, signifying that we effectively captured the trends in the data.

We chose our predictor variables to be the percent decrease in mobility (Mobility$_{t,i}$), median temperature (Temp$_{t,i}$), median humidity (Hum$_{t,i}$), and median PM$_{2.5}$ (PM$_{2.5_{t,i}}$). We also included the fixed effect of each province with the intercept $\lambda_i$ to account for local differences. Finally, we selected our response variable to be the log-transformed effective reproductive number ($R_{t,i}$). Choosing our response variable in this way allowed us to avoid lagging any of our predictor variables, as the effective reproductive number is already adjusted for the delay between the day of infection and the day the case was reported \citep{rtlive2020}. Furthermore, choosing the logarithm transformation allows the model to be interpreted more naturally as we can see the effects of the predictor variables on a relative scale rather than an absolute scale \citep{GelmanHill}.

The model was tested for robustness in two ways. First, we constructed the model for each province individually in order to ensure that the trends present in the global data set were also present in local data sets. Next, we cross-validated the model on our data by removing one province from the global data set, training the model on the 22 remaining provinces and using this model to predict $R_t$ for the removed province. We then repeated this process for each province in order to ensure that no province had an overstated effect on the model.

The GAM was computed using the \pkg{mgcv} package in \proglang{R} \citep{mgcv}.

\section{Results}%
\begin{figure}[t]
    \centering
    \includegraphics[width=\linewidth]{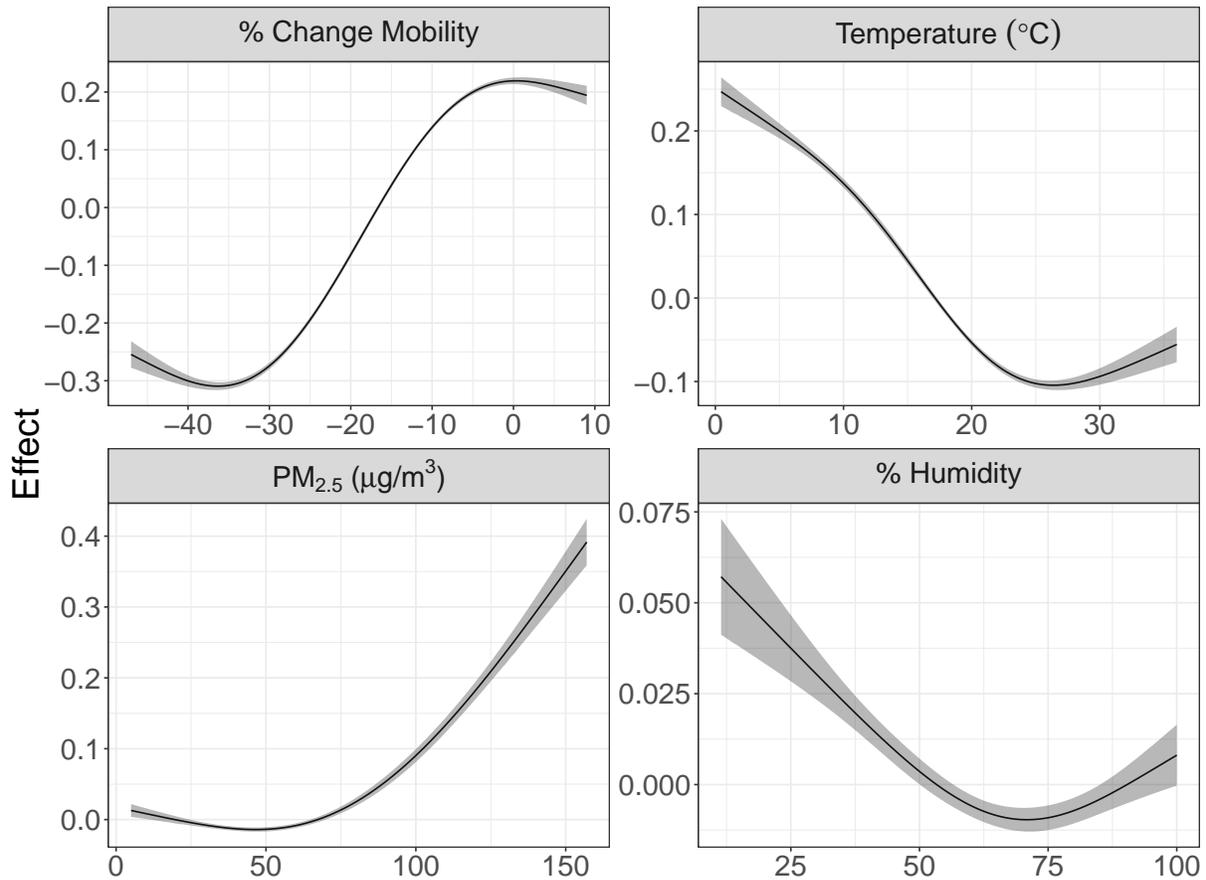}
    \caption{Each plot shows the partial effects of change in mobility, PM$_{2.5}$, temperature, and humidity on the expected value of $\log(R_t)$.  The 95\% confidence intervals are depicted in gray. $R_t$ increases as mobility increases, PM$_{2.5}$ has negligible effect for values between $0-70 \mu$g/m$^3$ but displays a strong positive effect at higher values, temperature has a negative relationship with $R_t$, and humidity has a slight positive effect under 50\% but a negligible effect at higher values.}
    \label{fig:GAM}
\end{figure}

\label{sec:Results}
Output from the GAM in Equation \eqref{eqn:gam} is shown in Figure \ref{fig:GAM}. These plots depict the partial effect of each predictor variable on the effective reproductive number $R_t$ determined by the \code{evaluate$\_$smooth} function in the \pkg{gratia} package in \proglang{R} \citep{gratia}. The 95\% confidence intervals are shown in gray. Negative values indicate a decrease in $R_t$, while positive values indicate an increase in $R_t$. If the confidence interval includes zero, there is no effect. 

The percent change in mobility had a positive effect on $R_t$. This is a very intuitive result; as people stay at home more, the effective reproductive number decreases.  Increased temperature showed a negative effect on the effective reproductive number up until around 25$\degree$C. Humidity had a somewhat positive effect when under 50$\%$ and a negligible effect at greater values.  PM$_{2.5}$ had no effect at concentrations between 0 and 70 $\mu$g/m$^3$, but showed a strong positive effect for larger values. All predictor variables were significant with p-values of less than $2 \times 10^{-16}$. The model had an adjusted coefficient of determination R$^2 = 0.67$.

\begin{figure}[t]
    \centering
    \includegraphics[width=\linewidth]{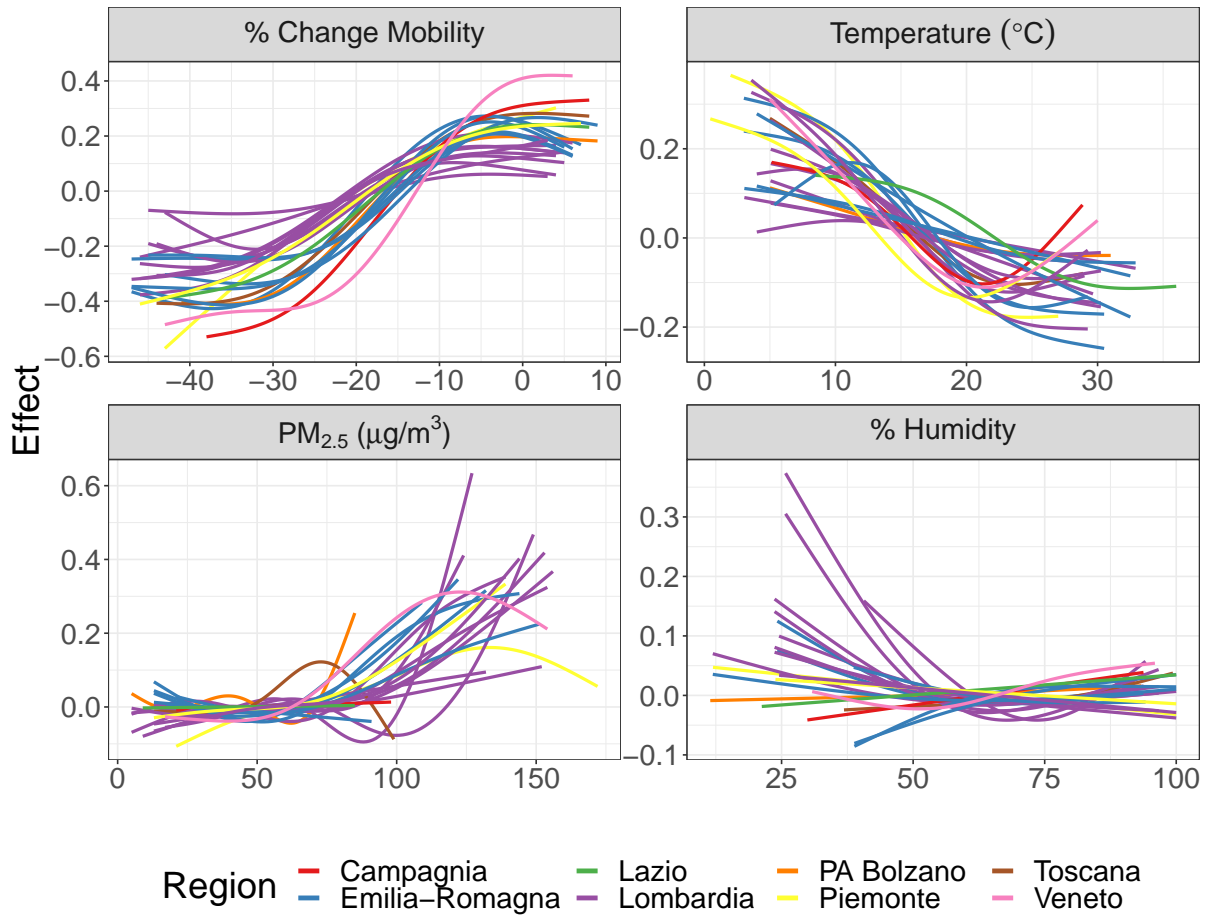}
    \caption{Each line shows the partial effect for each predictor variable when fitting a GAM to each province separately. Lines were colored by region which shows that the biggest outliers come from regions with only a small number of observations.}
    \label{fig:All25}
\end{figure}

The model was fit to each province individually which revealed similar qualitative effects for each predictor as observed in the global data set (Figure \ref{fig:All25}). Variability at higher PM$_{2.5}$ values exists due to a small number of observations per province. Firenze (Toscana), Napoli (Campagnia) and Roma (Lazio) do not exhibit the same increasing trend for PM$_{2.5}$ as the other provinces but all had very few observations above 75 $\mu \text{g/m}^3$ and no observations above 100 $\mu \text{g/m}^3$. Similarly, variability in the effect of humidity below 50\% can also be explained by a small sample size in this range per province.

\begin{figure}[t]
    \centering
    \includegraphics[scale=.2]{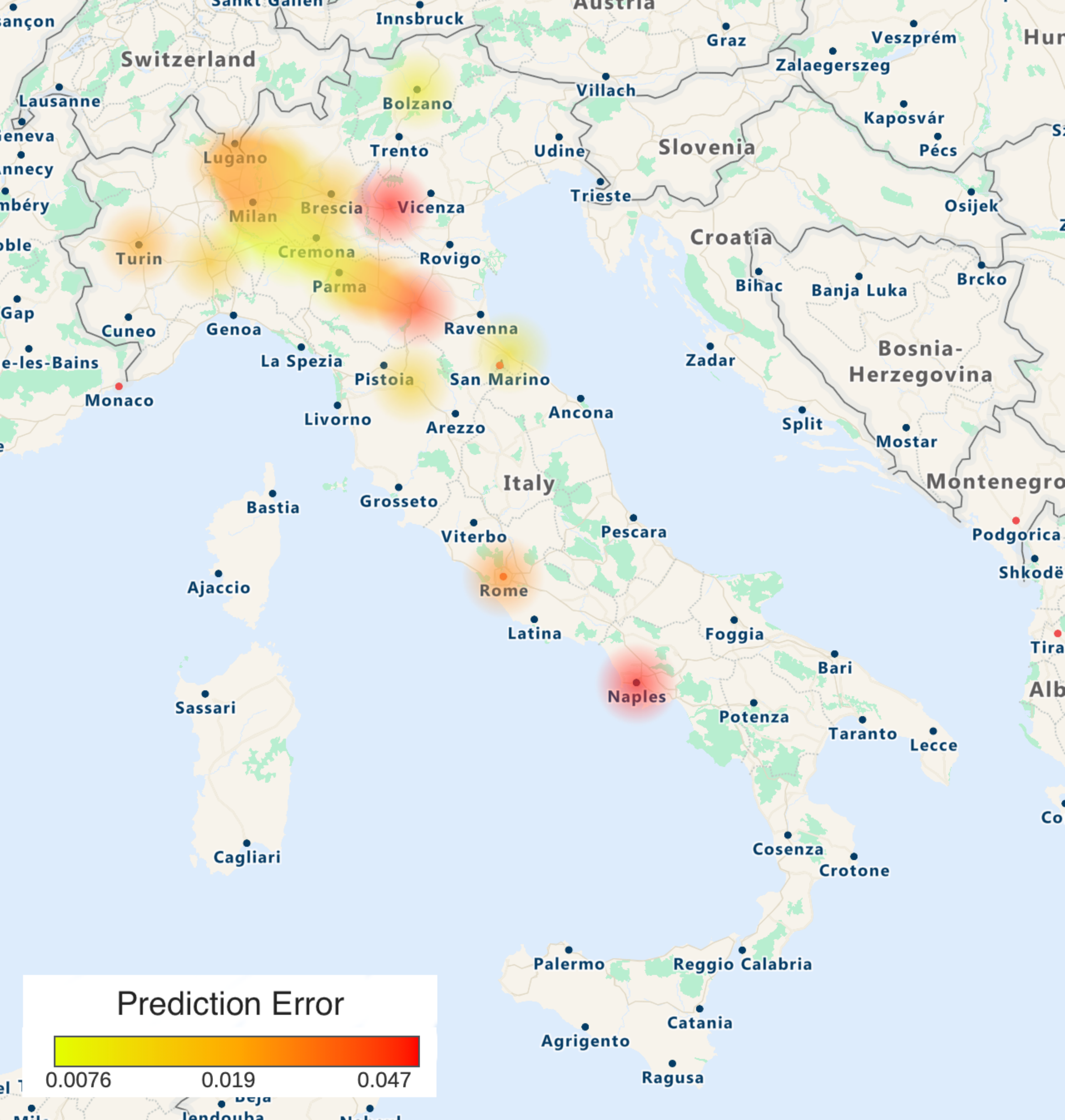}
    \caption{A geospatial heat map of mean squared prediction error from the cross validation process. The error is lowest in the Lombardi region, where we had the most observations, and highest in the regions of Campagni, Lazio, and Veneto where we used data from one province each.}
    \label{fig:error_map}
\end{figure}

Additionally, the model was trained on the data with one province removed at a time and the mean squared error was calculated after predicting on the removed province (Figure \ref{fig:error_map}). Prediction error was lower in regions where we had data from multiple provinces, and higher in regions where less data was available. Illustrating this, Napoli, Roma, and Verona were the only provinces in their respective regions from which we collected data, and they had the highest mean squared prediction errors.

\section{Discussion}%
\label{sec:Discussion}
The COVID-19 pandemic has affected millions of people around the world. Poor air quality is also considered a public health issue, associated with long-term health effects such as weakened cardiovascular health and decreasing life expectancy \citep{delfino2005potential, krewski2009evaluating, lelieveld2015contribution, janssen2013short}. Past studies have also shown an association between high concentrations of particulate matter less than 2.5 $\mu$m in diameter and an increase in influenza transmission \citep{su2019short}. Thus we hypothesize that even short-term exposure to elevated PM$_{2.5}$ may lead to a higher risk of COVID-19 transmission. Although any analysis to understand transmission is confounded by many variables, such as temperature, humidity, social distancing or the quality of data, our results show that air pollution may be associated with a higher effective reproduction rate of the virus. We showed the generalized additive model's results to be robust. When each province was removed from the data pool, a GAM constructed from the remaining provinces could predict case counts in the excluded province.

Our study is novel in its use of effective reproductive number as the dependent variable for the GAM. Although this parameter $R_t$ is commonly calculated to estimate the growth of a pandemic \citep{why_rt, why_rt2, why_rt3}, our use of it in a GAM analyzing the impact of other variables on disease spread is novel. As we were interested in a daily change in virus transmission, we chose to use the assumptions of an established model to account for the many unknowns in the data, such as the lag between contracting the virus and showing symptoms, and the one between taking a test and receiving a positive result. Furthermore, we use a longer timescale than comparable models of the COVID-19 pandemic, including both the initial wave of cases as well as the subsequent decline. Finally, we account for changes in human behavior during the pandemic using Google mobility data, while the majority of contemporary papers have not incorporated this or any similar metric.

We found statistically significant relationships of mobility, temperature, humidity, and PM$_{2.5}$ on COVID-19 transmission. It is well established that the reduction in mobility as a result of national lockdowns has been very effective at reducing the spread of COVID-19 \citep{effect_of_mobility, china_mobility_aqi_gam}, which is clear from our results as well. The magnitude of this effect explains why it was essential to include in our analysis. We also found an increase in temperature was associated with a decrease in $R_t$. We are hesitant to ascribe excessive significance to this finding given the strong serial correlation of temperature and the limited data we have available. Nevertheless, we can still compare our findings to those of other studies of COVID-19. \cite{166_countries} used a linear model on data from 166 countries to note a negative correlation between temperature and COVID-19 cases.  They also found a significant correlation between humidity and COVID-19 cases, albeit with a noticeably smaller effect than that of temperature. Both of these are supported by our results. Furthermore, \cite{temperature_covid_gam_brazil} used a non-linear model in Brazil to find a noticeable decrease in transmission as temperature increased from $17^{\circ}$C
to $24^{\circ}$C and little change between $24-27^{\circ}$C, which was similar to our findings. Studies of China such as those by \cite{temperature_covid_gam_china} and \cite{temperature_covid_gam_china2} have found the opposite result. They report positive nonlinear trends between temperature and disease incidence in January and February. Possible reasons for this disagreement may be model selection, confounding variables, or a limited range of data \citep{166_countries}.

Our results regarding PM$_{2.5}$ were also largely consistent with previous studies on COVID-19 and other respiratory illnesses. \cite{feng2016impact}, \cite{liang2014pm}, \cite{janssen2013short}, \cite{su2019short}, \cite{jiang2020effect}, \cite{yongjian2020association}, and \cite{china_mobility_aqi_gam} observed that PM$_{2.5}$ was positively correlated to the transmission of various viruses including coronaviruses and influenza viruses. Although most of these studies used linear models to establish a correlation, \cite{feng2016impact} reported a nonlinear trend with little correlation between infection rates for PM$_{2.5}$ values below 70 $\mu$g/m$^3$, followed by a linearly increasing trend for  PM$_{2.5}$ values between 70 and 150 $\mu$g/m$^3$, a similar phenomenon to the one we observed. Overall, it seems the majority of studies agree that a statistically significant relationship between PM$_{2.5}$ and viral transmission exists.  The common hypothesis for this is two-fold. First, fine particulate matter has been reported to damage the respiratory system, possibly resulting in ``a temporary immunosuppressive pulmonary microenvironment" \citep{feng2016impact, inhalation_toxicology, inhalation_toxicology2, inhalation_toxicology3, inhalation_toxicology4}. Second, PM$_{2.5}$ is sufficiently small that it can stay suspended in the air for prolonged stretches of time, making it, and any pathogen that may have bound itself to it, susceptible to inhalation. In fact, a recent finding from the Italian province of Bergamo provides evidence that SARS-CoV-2 does bind to particulate matter in the air \citep{bergamo_evidence}. Both of these hypotheses could imply that particulate matter allows greater exposure to the virus as well as a greater susceptibility to infection following exposure. However, more research is needed to understand the underlying mechanisms and the degree of additional risk. 

As pollution levels rise around the world \citep{shaddick2020half}, it is important to study their impact on respiratory disease. Even now, events such as wildfires in the western United States drive PM$_{2.5}$ levels to extreme highs \citep{liu2016particulate, xie2020summer}. Thus, our findings may predict an association between the fires and higher incidences of COVID-19 transmission. Furthermore, as the climate continues to change, this association may be important to note in future outbreaks of respiratory illnesses.

Another pervasive issue in the pandemic is the disproportional incidence of COVID-19 on low-income and ethnic communities, particularly in the United States \citep{racial_disparities, racial_disparities2, race_covid_air}.  Understanding the causes of this disparity would allow for policymakers to implement more effective measures of controlling the spread in these communities. To this end, it has been established that low-income and ethnic communities are disproportionately exposed to high levels of PM$_{2.5}$ \citep{race_covid_air, race_air}. The results of our study indicate that it is possible that some of the increased disease transmission in these communities may be attributed to higher levels of air pollution. Unfortunately due to a lack of available data, we were unable to directly investigate this using a subgroup analysis; however, we bring it up as an important issue that warrants further consideration.

In addition to PM$_{2.5}$, temperature, and humidity, it is possible to consider other environmental variables. Other studies have identified possible relationships between COVID-19 transmission and PM$_{10}$, SO$_2$, NO$_2$, and O$_3$ \citep{china_mobility_aqi_gam, jiang2020effect, Fronza2020}. We did not include these in our study for a variety of reasons. There were very few days where any province reported unhealthy NO$_2$ levels ($>$150 ppb), and as a result, we unsurprisingly found no strong effect. Likewise, O$_3$ measurements were missing from a significant number of provinces, and the O$_3$ that was reported seldom reached unhealthy ($>$100 ppb) levels. PM$_{10}$ and SO$_2$ measurements were similarly unavailable for a large number of provinces and thus were omitted from study. 

There are several limitations to our study. While we attempt to control for the seasonal variation of atmospheric variables in our model, an entire seasonal cycle of the COVID-19 pandemic has not been observed, making a full seasonal analysis infeasible at this time. Many confounding variables may influence PM$_{2.5}$ levels, and they are difficult to disentangle. Furthermore, since our study focuses on Italy, we do not know if these findings would be consistent in other areas. Future work would include expanding this study to multiple countries in different regions, such as the southern hemisphere, and to countries with a less homogeneous government response to test if our results hold on a global scale. 

\section{Conclusion}%
\label{sec:Conclusion}
Our results suggest that mobility, temperature, humidity, and short-term exposure to PM$_{2.5}$ are all possible risk factors for COVID-19. While the effects of temperature, humidity, and mobility have all been subjects of large-scale global studies on the pandemic \citep{effect_of_mobility, 166_countries}, PM$_{2.5}$ has received less attention. However, because PM$_{2.5}$ can change significantly due to both long-term, predictable events (such as an increase in PM$_{2.5}$ in winter months) as well as short-term, less predictable events (such as wildfires), it is important to try to understand how it may impact the pandemic. We find that PM$_{2.5}$ has a negligible correlation with $R_t$ for healthy to moderate air quality, but as PM levels increase to unhealthy amounts, a strong positive correlation emerges. This provides evidence for the long-standing hypothesis that short-term exposure to air pollution is a risk factor for respiratory illnesses. Although the results of our study are consistent with other works, this should still be considered a preliminary study as larger scale data sets are needed to confirm a global trend. 

\section*{Supplementary Materials}%
\label{sec:Supplementary}
The data and \proglang{R} code used to generate these results, as well as figures for individual cities in the validation process, are included in the supplementary files. A README is provided to describe the data available, how to generate each figure presented in the paper, and where important variables should be found in the code.

\section*{Acknowledgements}%
\label{sec:Acknowledgements}
We thank Dr. Richard McGehee, Dr. Mary Silber, and Dr. Mary Lou Zeeman for their contributions advising this effort as well as Dr. Jack O'Brien and Dr. Douglas Nychka for discussions regarding the development of the statistical model. This work was supported by the NSF, the American Institute of Mathematics, and the Mathematics and Climate Research Network. 

\bibliographystyle{jds}
\bibliography{bibliography}

\end{document}